\documentclass[aps,pra,twocolumn,reprint,showpacs,groupedaddress]{revtex4-1}
\usepackage{amssymb,amsmath,amsfonts} 
\usepackage{epsfig,graphicx}
\usepackage{latexsym}
\usepackage{stmaryrd}
\usepackage[table]{xcolor}
\usepackage[colorlinks=blue,citecolor=blue]{hyperref}
\usepackage{bm}
\usepackage{bbm}
\usepackage{color}
\usepackage[english]{babel}
\usepackage{subfigure}
\usepackage{natbib}
\usepackage{appendix}
\usepackage{amsmath}
\usepackage{graphicx}
\usepackage{amssymb}
\usepackage{txfonts}

\newcommand{\adag}{a^{\dagger}}
\newcommand{\adaga}{a^{\dagger}a}
\newcommand\ket[1]{\left|\textstyle{#1}\right\rangle}

\newcommand{\upx}{\ket{\uparrow}_x}

\makeatletter

\@ifundefined{textcolor}{}
{%
 \definecolor{BLACK}{gray}{0}
 \definecolor{WHITE}{gray}{1}
 \definecolor{RED}{rgb}{1,0,0}
 \definecolor{GREEN}{rgb}{0,1,0}
 \definecolor{BLUE}{rgb}{0,0,1}
 \definecolor{CYAN}{cmyk}{1,0,0,0}
 \definecolor{MAGENTA}{cmyk}{0,1,0,0}
 \definecolor{YELLOW}{cmyk}{0,0,1,0}
}

\makeatother

\begin{document}
\title{Protected ultrastrong coupling regime of the two-photon quantum Rabi model with trapped ions}
\author{Ricardo Puebla, Myung-Joong Hwang, Jorge Casanova and Martin B. Plenio}
\affiliation{Institut f\"{u}r Theoretische Physik and IQST, Albert-Einstein-Allee 11, Universit\"{a}t Ulm, D-89069 Ulm, Germany}

\begin{abstract}
We propose a robust realization of the two-photon quantum Rabi model in a trapped-ion setting based on a continuous dynamical decoupling scheme. In this manner the magnetic dephasing noise, which is identified as the main obstacle to achieve long time coherent dynamics in ion-trap simulators, can be safely eliminated. More specifically, we investigate the ultrastrong coupling regime of the two-photon quantum Rabi model whose realization in trapped ions involves second-order sideband processes. Hence, the resulting dynamics becomes unavoidably slow and more exposed to magnetic noise requiring an appropriate scheme for its elimination. Furthermore, we discuss how dynamical decoupling methods take a dual role in our protocol, namely they remove the main source of decoherence from the dynamics while actively define the parameter regime of the simulated model.
\end{abstract}


\maketitle

\section{Introduction} \label{sec:intro}
Dealing with almost perfectly isolated quantum registers is nowadays possible thanks to the development of modern quantum technologies~\cite{Monroe02, Dowling02}. However, inherent experimental imperfections still challenge the achievement of isolated quantum systems leading to decoherence~\cite{Nielsen00} and, therefore, to the loss of the quantum properties of the system. Although the impact of these imperfections might be reduced by upcoming technological advances, the use of schemes of external control that provides with robustness against noise is preferred as they are currently feasible and potentially adaptable to different quantum platforms. In this respect, decoherence free subspaces~\cite{Lidar12}, quantum error correction~\cite{Lidar13}, and dynamical decoupling techniques~\cite{Souza12} find themselves amongst the most employed strategies to cope with different sources of noise as a result of their simplicity and versatility. 

On the other hand, trapped-ions represent one of the most promising platforms for quantum information processing due the combination of long coherence times for the qubits with the ability to initialize the registers, and to perform high-fidelity coherent operations and measurements~\cite{Leibfried03, Haffner08, Leibfried03bis, Roos06, Myerson08}. In particular, while different setups may suffer from distinct noise sources, quantum coherence in typical experiments involving metastable states of optical ions or  microwave driven ions is mainly limited by fluctuating magnetic fields which leads to dephasing noise. In addition, thermal noise and imperfections in the applied-radiation parameters, to be commented later, might be also of importance~\cite{Leibfried03, Haffner08}. It is then desirable to develop methods to cope with magnetic-dephasing noise; see for example the techniques in~\cite{Timoney11, Tan13, Weidt16} that are designed to achieve noise resilient two-qubit gates. These methods, besides being feasible to implement, ought to extend the quantum coherence of the system by removing different noise contributions  and provide with a wide range of parameter regimes  to display the aimed quantum evolution. 

In this article we propose a scheme based on continuous dynamical decoupling techniques to carry out a robust realization of the two-photon quantum Rabi model  (2PQRM) in a trapped ion. This is achieved via the application of external radiation (laser) that plays multiple roles in our scheme; they couple the internal degrees of freedom of a single trapped-ion with the external (vibrational) ones, determine the range of simulated parameter regime, and suppress magnetic fluctuations in the dynamics. A trapped-ion simulation of the 2PQRM has been recently proposed in Ref.~\cite{Felicetti15} as a natural extension of the quantum Rabi model (QRM) simulation. The latter in its different forms, e.g.~\cite{Pedernales15, Puebla16}, has served to study intriguing phenomena of the QRM such as the emergence of a quantum phase transition~\cite{Hwang15,Puebla17}, or the appearance of collapses and revivals of the qubit population in the deep strong coupling regime~\cite{Casanova10}. The 2PQRM features interesting physics especially in its ultrastrong coupling (USC) regime, which we define in analogy to the USC regime in the QRM~\cite{Niemczyk10,FornDiaz10}, namely, when the coupling constant becomes comparable to the bosonic frequency. However, because the two-photon interaction terms in the 2PQRM is realized through slow second-order sideband processes, the simulation of the dynamics in the USC regime is unavoidably slow and is therefore prone to noise that spoils its realization. Therefore, a robust scheme against magnetic fluctuations may become inevitable for a faithful realization of the dynamics in the USC regime.

This article is organized as follow. The Sec.~\ref{sec:2PQRM} is dedicated to introduce the 2PQRM, while in Sec.~\ref{sec:TI} we discuss the applicability of dynamical decoupling methods in the context of optical trapped-ions. In the Sec.~\ref{sec:NS} it is demonstrated, by means of detailed numerical simulations, that our protected trapped-ion scheme enhances the completion of the 2PQRM with respect to the bare case where the magnetic-dephasing noise damages its realization. As a result, a faithful and robust simulation of the 2PQRM is accomplished. Finally, in Sec.~\ref{sec:conc} we summarize the main results.

\section{Two-photon quantum Rabi model}
\label{sec:2PQRM}

The 2PQRM describes the interaction between a single two-level system and a bosonic mode. However, the coupling term differs from that of the QRM because the interaction is based on the exchange of two bosonic excitations. The Hamiltonian of the 2PQRM reads  $(\hbar=1)$
\begin{align}
\label{eq:2PQRM}
H_{\rm {2PQRM}}=\frac{\tilde{\Omega}}{2}\sigma_z+\tilde{\omega}_0\adaga +\tilde{g}\left[a^2+(\adag)^2\right]\sigma_x.
\end{align}
The Pauli operators $\sigma_{x,y,z}$ act on the two-level system, whose frequency splitting is $\tilde{\Omega}$. Additionally, $a$ and $\adag$ represent the usual annihilation and creation operators of the quantized bosonic mode with a frequency $\tilde{\omega}_0$, and a corresponding coupling parameter $\tilde{g}$ between both subsystems. For the sake of clarity, parameters with tilde will refer throughout the article to those of the $H_{\rm {2PQRM}}$.

The  2PQRM has been acknowledged to play a role in preparing non-classical states of light and to correctly describe second-order processes across different systems~\cite{Brune87,Toor92, Bertet02, Stufler06, delValle11}. Furthermore, the 2PQRM exhibits some interesting features which have been recently subject to study, such as its integrability or the appearance of spectral collapse~\cite{Chen12, Peng13, Felicetti15, Duan16}. Indeed, regardless of the specific value of $\tilde{\Omega}$, when the coupling constant reaches the value $\tilde{g}_{sc}=\tilde{\omega}_0/2$, the spectrum becomes continuous above a certain excitation energy and the bosonic population diverges in excited eigenstates~\cite{Felicetti15,Duan16}. In addition, $H_{\rm 2PQRM}$ becomes unbounded from below for larger couplings, $\tilde{g}>\tilde{g}_{sc}$. Certainly, these phenomena take place in the USC regime as $\tilde{g}$ becomes a large fraction of the bosonic frequency $\tilde{\omega}_0$. This regime can be achieved in a trapped-ion experiment as recently proposed in Ref.~\cite{Felicetti15} where the coupling parameter can be brought close to $\tilde{g}_{sc}=\tilde{\omega}_0/2$. 

However, as we will see later, a major constraint to the exploration of the USC regime for two-photon processes in a trapped-ion setting resides in the relatively small couplings that can be accomplished. As a consequence, the desired dynamics takes place in a time scale that may exceed the coherence time in the trapped-ion setup, and thus, it will be spoiled by the presence of realistic experimental imperfections, such as magnetic-field fluctuations. Hence, a prolonged coherence time or a noise-resilient scheme turn out to be crucial to explore this regime.   In the following we propose a scheme to faithfully realize the 2PQRM based on continuous dynamical decoupling to prolong the coherence time of the system, originally limited by the presence of dephasing noise.

\section{2PQRM realization with a trapped ion}
\label{sec:TI}

In this section we present first the standard or bare scheme for achieving a 2PQRM using a trapped ion. Then, in Sec.~\ref{sub:noise} we describe the magnetic-field fluctuations, their incorporation to the trapped-ion dynamics, and how these fluctuations certainly spoil the bare realization of the 2PQRM. However, they can be overcomed by means of a continuous dynamical decoupling scheme, which is explained in Sec.~\ref{sub:dd} together with its actual trapped-ion implementation.

Consider a single atomic ion where two particular internal electronic states, separated by a corresponding frequency splitting $\omega_I$ define a qubit, and it is placed in a trap of frequency $\nu$.   Then, the corresponding Hamiltonian of the trapped ion and its interaction with different irradiation sources, characterized by Rabi frequency $\Omega_j$, wave vector $\vec{k}_j$ with $k_j$ its component in the direction of the ion's vibration, frequency $\omega_j$, and initial phase $\phi_j$, can be written as~\cite{Leibfried03, Haffner08}
\begin{equation}
\label{eq:HTI}
H_{TI}= \frac{\omega_I}{2}\sigma_z+\nu\adaga+\sum_j\frac{\Omega_j}{2}\sigma_x\left [e^{i(k_j\hat{x}-\omega_j t-\phi_j)}+{\rm H.c.}\right].
\end{equation}
Here,  $\hat{x}$ stands for the ion position operator in the harmonic trap, $\hat{x}=(2m\nu)^{-1/2}\left(a+\adag\right)$, with $m$ the ion mass. It is useful to introduce the so-called Lamb-Dicke parameter $\eta_j=k_j(2m\nu)^{-1/2}$ which quantifies the coupling between the internal qubit states and the motional states of the ion. To better understand the forthcoming approximations, it is helpful to have in mind typical values for trapped-ion experiments. In particular, we consider those of an optical ion, ${}^{40}{\rm Ca}^{+}$, which are $\nu=2\pi \times 2.0$ MHz and $\omega_I=2\pi \times 4\cdot 10^{14}$ Hz, that corresponds to an optical wavelength of $729$ nm. The corresponding Rabi frequencies of the lasers lie in the range of kHz, while the Lamb-Dicke parameter is typically $\eta\sim 10^{-2}$~\cite{Gerritsma10, Gerritsma11}. We will focus in this specific system through the rest of the article.

As a consequence of the considered parameters, the optical rotating wave approximation (RWA) can be safely invoked~\cite{Leibfried03, Haffner08}, i.e., terms oscillating at frequency $\omega_I+\omega_j$ can be neglected (note that $\omega_I+\omega_j \gg \Omega_j$), while keeping those that evolve at a speed of $\omega_I-\omega_j$. 
Then, in an interaction picture with respect to the free Hamiltonian $H_0=\frac{\omega_I}{2} \sigma_z+\nu\adaga$, and within the Lamb-Dicke regime, $\eta \sqrt{ \langle (a+a^\dag)^2 \rangle}\ll 1$, the previous Hamiltonian can be further developed to give rise to Jaynes-Cummings and/or anti Jaynes-Cummings structures~\cite{Haffner08}. In order to achieve the two-photon coupling, two driving lasers tuned to $\omega_{r,b}=\omega_I \mp 2\nu-\delta_{r,b}$ are needed. In the latter equation the subscripts $r$ and $b$ denote the lasers that drive second order red- and blue-sidebands respectively~\cite{Leibfried03, Felicetti15}. The detunings $\delta_{r,b}$ are chosen to be small, $\left|\delta_{r,b}\right|\ll \nu$ which together with the condition $\Omega_{r,b}\ll \nu$, allow us to apply the vibrational RWA, and thus, the trapped-ion Hamiltonian adopts the following form
\begin{align}
\label{eq:TIa}
H_{TI}^{I}\approx -\frac{\Omega_r}{4}&\eta_r^2\left(\sigma^+a^2e^{i\delta_r t}e^{-i\phi_r}+{\rm H.c.}\right)\nonumber\\
&-\frac{\Omega_b}{4}\eta_b^2\left(\sigma^+(\adag)^2e^{i\delta_b t}e^{-i\phi_b}+{\rm H.c.}\right).
\end{align}
Choosing  $\phi_r=\phi_b=\pi$, $\Omega_{r}=\Omega_{b}=\Omega$ and $\eta_r=\eta_b=\eta$, the previous Hamiltonian acquires the form of a 2PQRM $H_{\rm 2PQRM}=\tilde{\Omega}/2 \sigma_z+\tilde{\omega}_0\adaga+\tilde{g}\left[a^2+(\adag)^2\right]\sigma_x$ in an interaction picture with respect to $H_0=\tilde{\Omega}/2 \sigma_z+\tilde{\omega_0}\adaga$, where $\tilde{\Omega}= (\delta_b+\delta_r)/2$, $\tilde{\omega}_0=(\delta_b-\delta_r)/4$ and $\tilde{g}=\eta^2\Omega/4$. This corresponds to a realization of the 2PQRM with a trapped ion~\cite{Felicetti15}.  Nevertheless, beyond the correct functioning of the aforementioned approximations, magnetic field fluctuations may set a tighter constraint to the actual realization of the model, which  have not been introduced so far. As anticipated, as a consequence of small coupling,  $\tilde{g}\propto \eta^2$, a simulation time for the strong and ultrastrong coupling regime dynamics becomes long. Note that  $\tilde{g} \sim 10^2$ Hz and consequently, to achieve USC, $\tilde{\omega}_0  \sim 10^2$ Hz. As a rough estimate, relevant effects of the Hamiltonian will take place when $\tilde{g}t\sim 1/2$, which leads to simulation times, $t\sim 5$ ms, already of the order of the coherence time $T_2$ (see following subsection).  It is therefore desirable to attain, by means of a simple scheme, a longer and noise-resilient, yet tunable, realization of the 2PQRM, which is precisely the objective of the present article.

\subsection{Magnetic-dephasing noise}
\label{sub:noise}
Typical experiments performed with the metastable internal states of optical ions, as it is the case for the ${}^{40}{\rm Ca}^{+}$ ion, show magnetic fluctuations which set a coherence time of the qubit  to $T_2\approx 3$ ms~\cite{Gerritsma11}, while noises such as phonon heating and qubit decay (with rates $\sim 10$ Hz and $\sim 1$ Hz respectively) are expected to play a role in a much longer time scale~\cite{Schmidt-Kaler03}.  In the same manner, trapped-ion experiments involving magnetic field gradients are also affected by magnetic-field fluctuations exhibiting similar coherence times, as in the case of the microwave driven ${}^{171}{\rm Yb}^+$ ion~\cite{Timoney11}. Therefore, no further sources of noise have been included  in our analysis since the examined total evolution time is, at least, one order of magnitude shorter than their typical rates. It is then expected that their effect is negligible. 

The effect of magnetic-field fluctuations in the trapped-ion Hamiltonian can be effectively captured by adding an extra stochastic $\sigma_z$ term. Therefore, the noisy trapped-ion Hamiltonian becomes $H_{TI,n}=H_{TI}+H_{dp}$, where $H_{dp}$ simply reads
\begin{align}
\label{eq:dp}
H_{dp}=\frac{\xi(t)}{2}\sigma_z.
\end{align}
This magnetic-field fluctuation can be modeled as an Orstein-Uhlenbeck process~\cite{Uhlenbeck30, Wang45, Gillespie96, Gillespie96bis}, as illustrated in~\cite{Bermudez12,Lemmer13,Puebla16}, since it successfully describes the decay of coherence due to dephasing noise~\cite{Wineland98, Cai12}.   

The time evolution of the stochastic variable $\xi(t)$ within the Orstein-Uhlenbeck model  depends only on two variables, namely, correlation time $\tau$ and diffusion constant $c$,
\begin{align}
\label{eq:OU}
&\xi(t)=\xi(0)e^{-t/\tau}+\left[\frac{c\tau}{2}\left(1-e^{-2t/\tau}\right) \right]^{1/2}N(t), \\
&\qquad \quad \overline{N(t)}=0, \quad \overline{N(t)N(t')}=\delta(t-t'),
\end{align}
where the overline denotes stochastic average and $N(t)$ represents a normal-distributed random variable. While the correlation time $\tau$ determines the width of the spectral density, the diffusion constant is proportional to the total power. That is, $\tau$ defines a characteristic frequency, $f_{cr}=1/(2\pi\tau)$, at which the spectral density changes from $S(f< f_{cr})\propto f^0$ to $S(f> f_{cr})\propto f^{-2}$. Accordingly, both values, $c$ and $\tau$,  are linked to the coherence time $T_2$ of the system, $T_2\approx 2/(c \tau^2)$ for $\tau\ll T_2$~\cite{Gillespie96,Gillespie96bis}. This is indeed the typical scenario in trapped ions, where the quantum coherence decays exponentially in a time longer than the correlation time of magnetic-field fluctuations~\cite{Wineland98}. In Ref.~\cite{Mikelsons15}, based on trapped-ion measurements, $\tau=100 \ \mu$s is proposed as a good estimate for the correlation time of the magnetic noise. Hence, in the following we will consider  $\tau=100 \ \mu$s and a coherence time $T_2=3$ ms. We refer to the Appendix~\ref{app:a} for further details.

\subsection{Protected 2PQRM: Continuous dynamical decoupling}
\label{sub:dd}
In this part we summarize how to overcome the effect of dephasing noise by continuous dynamical decoupling, and how to implement it in a trapped-ion setup for a robust realization of the 2PQRM. In short, this method consists in using  an additional driving field that gives rise to a dressed basis in which the effect of the noise is suppressed. This is achieved if the energy gap in the new dressed basis is sufficiently large such that noise is not capable of producing transitions due to energy conservation~\cite{Puebla16, Cai12}. Continuous dynamical decoupling is useful in providing protection when a system is subject to a time-correlated noise, that is, a noise with a finite spectral width. This is indeed the case of the magnetic-dephasing noise, which can be modeled as described earlier. For the case of pulsed dynamical decoupling, see for example~\cite{Souza12,Casanova15}.

The simplest case consists in considering the Hamiltonian $H= \xi(t)/2\sigma_z + \omega_0/2\sigma_z $, i.e. one qubit with an energy splitting $\omega_0$ that it is affected by the noise $\xi(t)$. Then, if a resonant driving field is introduced in the system, after the RWA and in a suitable interaction picture the dynamics is described by 
\begin{align}\label{simplest}
H=\frac{\xi(t)}{2}\sigma_z+\frac{\Omega_{DD}}{2}\sigma_x,
\end{align}
where $\Omega_{DD}$ corresponds to the intensity of the dynamical decoupling driving field. This Hamiltonian can be approximated by the noiseless Hamiltonian $H\approx \Omega_{DD}/2\sigma_x$ if the time-dependent $\xi(t)$ is not too strong and has vanishingly small Fourier components at frequency $\Omega_{DD}$. More precisely, let  $\hat{\xi}(f)=\int dt \ \xi(t) e^{i 2\pi f t}$ be the Fourier transform of $\xi(t)$, then, in a rotating frame with respect to $\Omega_{DD}/2\sigma_x$ the Hamiltonian~(\ref{simplest}) becomes approximately the identity if $|\hat{\xi}(f)|\ll |\Omega_{DD}\pm f|$, as a result of applying the RWA. 

In our effective description of magnetic-field fluctuations $\xi(t)$  corresponds to an Orstein-Uhlenbeck process, and hence, the basic requirement for noise suppression is $\Omega_{DD}> f_{cr}$  (see Appendix~\ref{app:a} for further details). Finally, we would like to comment that the presented scheme allows for a concatenated scheme~\cite{Cai12}, in which further sources of noise can be handled introducing consecutive driving fields, as shown recently in~\cite{Puebla16}.

In the following we show how to implement a two-photon Rabi model with protection against magnetic-dephasing noise based on the aforementioned continuous dynamical decoupling technique. Besides the  lasers that give rise to the two-photon (phonon) Hamiltonan, in our scheme dynamical decoupling is achieved by introducing an additional laser driving a \textit{carrier} interaction (denoted with a subscript $c$), i.e. the laser frequency $\omega_c$ is tuned as $\omega_c=\omega_I$, with its corresponding Rabi frequency and Lamb-Dicke parameter denoted by $\Omega_c$ and $\eta_c$, respectively.  In this manner, in a rotating frame with  respect to $H_0$, and considering  optical RWA, Lamb-Dicke regime and vibrational RWA,  the trapped-ion Hamiltonian reads
\begin{align}
H_{TI}^{I}\approx \frac{\xi(t)}{2}\sigma_z&+\frac{\Omega_c}{2}\left(\sigma^+e^{-i\phi_c}+\sigma^-e^{i\phi_c} \right) \nonumber \\&-\frac{\Omega_r}{4}\eta_r^2\left(\sigma^+a^2e^{i\delta_r t}e^{-i\phi_r}+{\rm H.c.}\right)\nonumber \\& -\frac{\Omega_b}{4}\eta_b^2\left(\sigma^+(\adag)^2e^{i\delta_b t}e^{-i\phi_b}+{\rm H.c.}\right),
\end{align}
where we have set the frequencies of the sidebands to $\omega_{r,b}=\omega_I\mp 2\nu -\delta_{r,b}$. Then, choosing an initial phase for the carrier driving $\phi_c=0$, we obtain a new free energy term for the qubit, $\Omega_c/2\sigma_x$, which is orthogonal to the magnetic-dephasing noise.

Note however that if $\Omega_c$ is directly used to define the new dressed basis to simulate the 2PQRM, i.e. $\Omega_c\equiv \tilde{\Omega}$ in Eq.~(\ref{eq:2PQRM}), the accessible simulated parameters are constrained to the cases where  $\tilde{\Omega}/\tilde{\omega}_0\gg 1$, as $\Omega_c $ must be large enough to ensure decoupling, i.e. $\Omega_{c}> f_{cr}\sim$ kHz for $\tau=100 \ \mu$s, and because $\tilde{\omega}_0 \simeq \tilde{g}\sim 10^2$ Hz to explore USC regime.

Therefore, in order to avoid this unnecessary restriction which reduces the tunability of the protected 2PQRM, we make use of a more general scheme which grants a wide tunability of simulated parameters. More specifically, the intensity of the carrier laser can be divided in two parts, $\Omega_c=\Omega_{DD}+\tilde{\Omega}$. Then, we move to an additional interaction picture with respect to $\Omega_{DD}/2\sigma_x$, where it can be seen that  magnetic-field fluctuations are largely suppressed when $\Omega_{DD}> f_{cr}$, while leaving  $\tilde{\Omega}$ as a tunable free parameter. Therefore, the driven carrier takes a double role in our protocol, i.e. it allows to average out magnetic-field fluctuations, while defines $\tilde{\Omega}$, the desired energy gap of the two-level system.   Choosing $\eta_r=\eta_b=\eta$ and $\Omega_r=\Omega_b=\Omega$, the Hamiltonian reads
\begin{align}
H_{TI}^{II}&\approx \frac{\tilde{\Omega}}{2}\sigma_x \nonumber \\
&-\frac{\eta^2 \Omega}{8} \bigg\{ \left[ \sigma_x+i\left( \cos(\Omega_{DD}t) \sigma_y-\sin(\Omega_{DD}t)\sigma_z\right) \right]  \nonumber \\ & \left[a^2e^{i\delta_r t}e^{-i\phi_r}+(\adag)^2e^{i\delta_b t}e^{-i\phi_b}\right] +{\rm H.c.} \bigg\}.
\end{align}
Finally, we require that the detunings of  second sidebands are $\delta_{r,b}=\Omega_{DD} \mp 2\tilde{\omega}_0$, and then we invoke an additional RWA to neglect terms rotating at frequencies $\pm \Omega_{DD}$ and higher, which leads to
\begin{align}
\label{eq:TIb}
H_{TI}^{II}\approx \frac{\tilde{\Omega}}{2}\sigma_x +\frac{\eta^2 \Omega}{8}\sigma_z\left[a^2e^{-2i\tilde{\omega}_0 t}+ (\adag)^2e^{2i\tilde{\omega}_0 t} \right],
\end{align}
where we have set already $\phi_r=\phi_b=\pi$. The previous Hamiltonian corresponds to the 2PQRM in Eq.~(\ref{eq:2PQRM}), in the rotating frame of $\tilde{\omega}_0\adaga$,  where the spin basis has been rotated. In addition we have that the coupling constant $\tilde{g}=\eta^2 \Omega/8$, is half of the achieved coupling that appears in the case without protection, see Eq.~(\ref{eq:TIa}). In this manner, the evolution time for the protected case has to be doubled to observe the same dynamics than in the bare realization, nevertheless dynamical decoupling enables a faithful realization during much longer times, as we demonstrate in Sec.~\ref{sec:NS}.

 Note that to attain Eq.~(\ref{eq:TIb}) one requires the conditions $\Omega_{DD}\gg \tilde{\omega}_0$ and $\Omega_{DD}\gg \eta^2\Omega/8$. Certainly, a large intensity $\Omega_{DD}$ works in favor of the previous RWA and to better decouple the system from the addressed fluctuations. However, trap frequency sets a fundamental limitation since second sidebands are driven, and thus, $|\delta_{r,b}|\ll \nu$ or equivalently, the intensity $\Omega_{DD}$ must be small enough when compared with the trap frequency, $\Omega_{DD}\ll \nu$. To the contrary, we emphasize that $\tilde{\Omega}$ is a free parameter only limited by the intensity of the driven carrier, whose typical value does not restrict the validity of any approximation to achieve the desired 2PQRM, Eq.~(\ref{eq:TIb}). Thus, the proposed scheme allows to realize the 2PQRM in far detuned scenarios,  $\tilde{\Omega}\gg \tilde{\omega}_0$ or $\tilde{\Omega}\ll \tilde{\omega}_0$, or in near resonant condition, $\tilde{\Omega}\approx \tilde{\omega}_0$.
In this respect, typical trapped-ion parameters provide enough room to faithfully implement the 2PQRM in different regimes with  simultaneous noise elimination as shown by means of numerical simulations in the following section.

\section{Numerical simulations}
\label{sec:NS}
\begin{figure}
\includegraphics[width=1.2\linewidth,angle=-90]{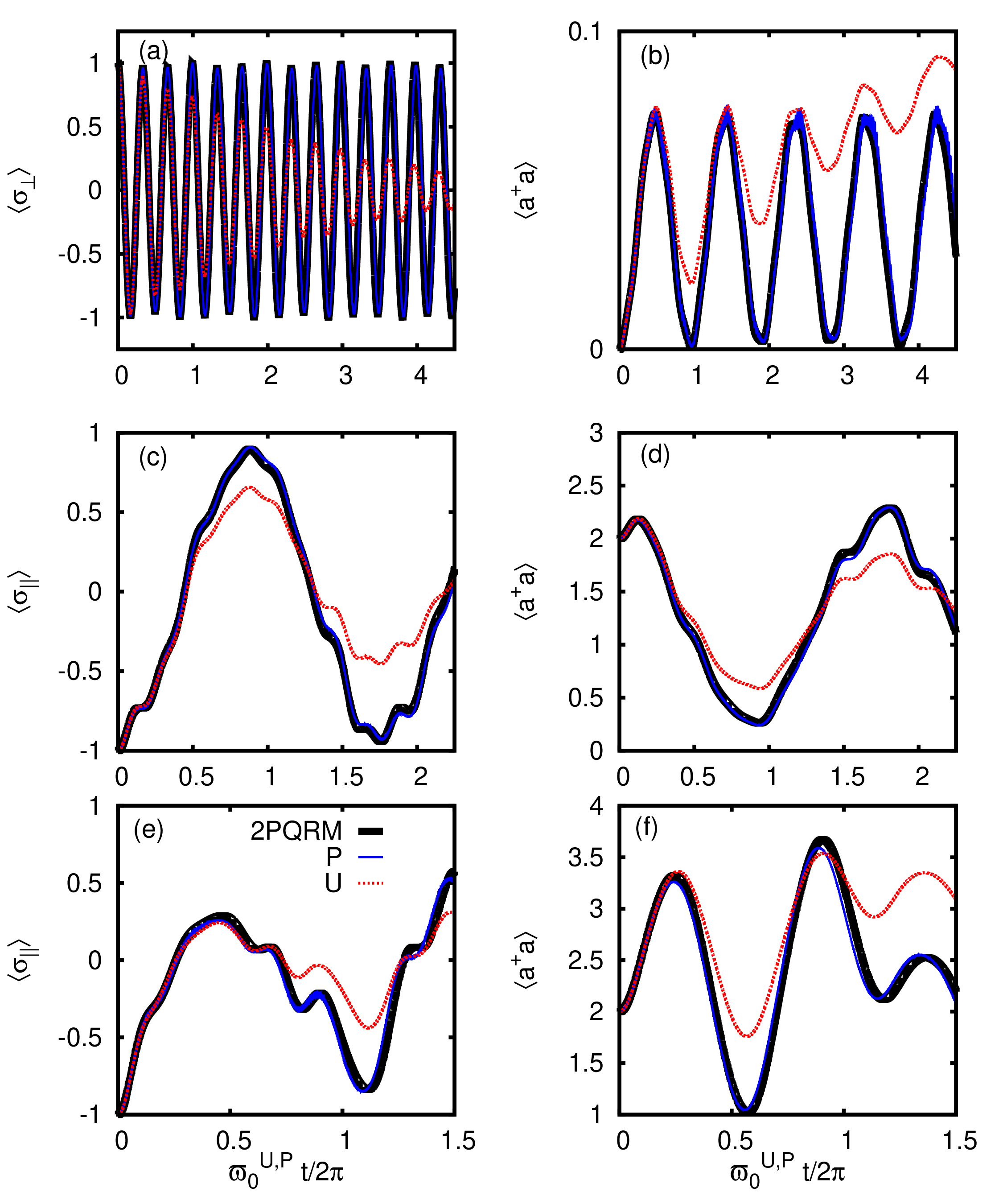}
\caption{(Color online) Dynamics of the trapped-ion realization of the 2PQRM, with protection (P) (blue solid line) and without (U) (red dotted line), together with the ideal noiseless 2PQRM (black solid line). In the left column qubit observables, and $\left<\adaga\right>$ in the right. The results were obtained averaging $400$ stochastic trajectories. Three different scenarios are examined. Panels (a) and (b) correspond to $\tilde{g}/\tilde{\omega}_0=0.1$, $\tilde{\Omega}/\tilde{\omega}_0=3$ and initial state $\left|\downarrow \right>_\perp\left|0 \right>$. Panels (c) and (d) to $\tilde{g}/\tilde{\omega}_0=0.2$, $\tilde{\Omega}/\tilde{\omega}_0=2$ and initial state $\left|\uparrow \right>_\parallel\left|2 \right>$, and at the bottom, (e) and (f), $\tilde{g}/\tilde{\omega}_0=0.3$, $\tilde{\Omega}/\tilde{\omega}_0=1$ and initial state $\left|\downarrow \right>_\parallel\left|2 \right>$. The total time for the unprotected realization is $5$ ms, and the corresponding double time for the protected scheme, $10$ ms. See Tables~\ref{tab:1} and~\ref{tab:2}, as well as the main text, for further details. Note the meaningful improvement in the simulated dynamics using continuous dynamical decoupling, which basically lies on top of the noiseless 2PQRM.}
\label{fig:1}
\end{figure}
\begin{figure}
\includegraphics[width=0.5\linewidth,angle=-90]{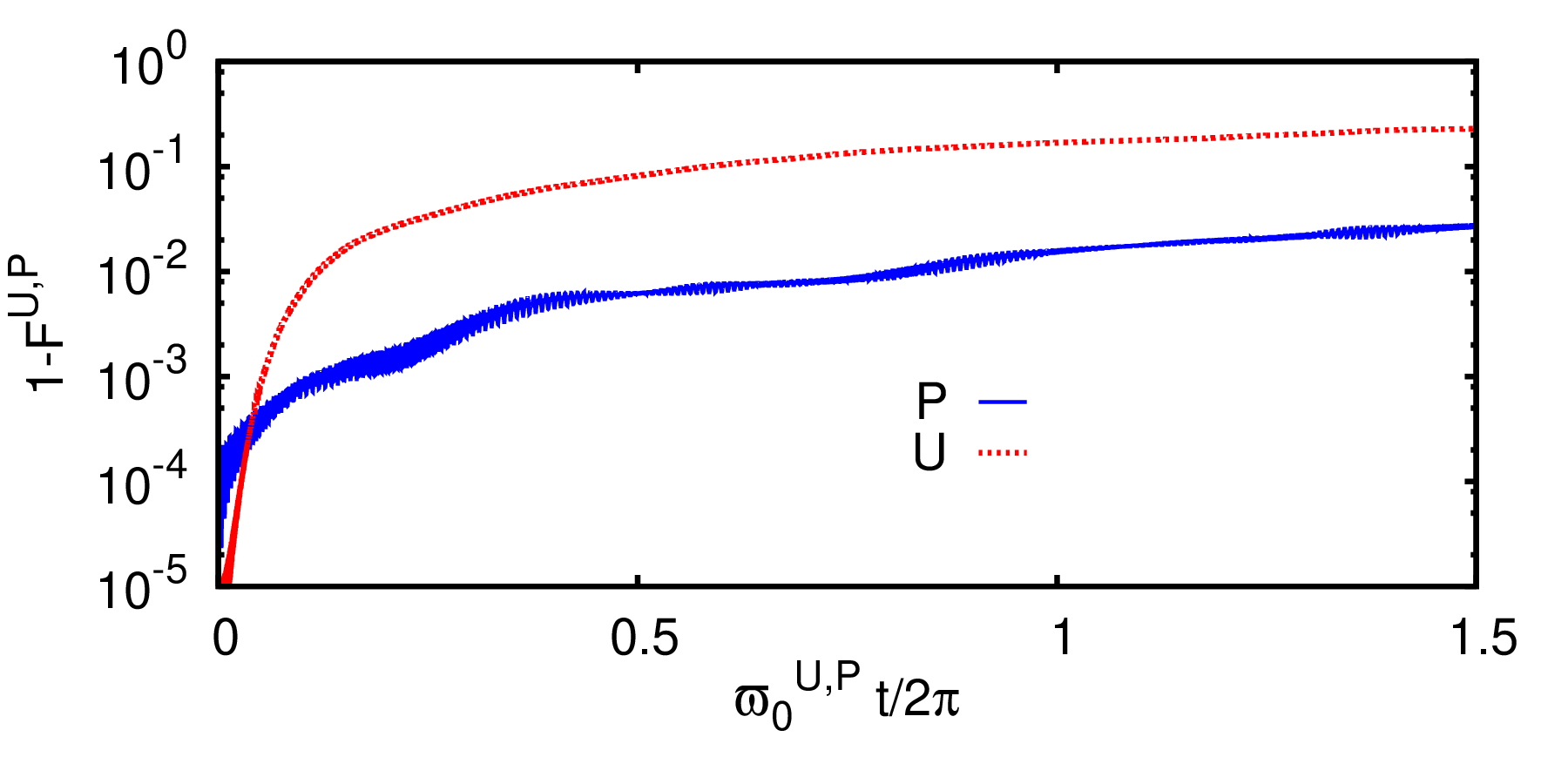}
\caption{ (Color online) Infidelity, $1-F^{U,P}$, between the initial state $\left|\downarrow \right>_\parallel\left|2 \right>$  evolving under the unprotected (red dotted line) and protected (blue solid line) trapped-ion Hamiltonian realization and the targeted ideal 2PQRM, with simulated parameters $\tilde{g}/\tilde{\omega}_0=0.3$ and $\tilde{\Omega}=\tilde{\omega}_0$. The results correspond to an ensemble average over $400$ stochastic trajectories. Note the substantial improvement of one order of magnitude in the infidelity respect to the unprotected scheme, whose total time amounts to $5$ ms; double time for the protected version.}
\label{fig:2}
\end{figure}
In this part we show the results of the trapped-ion numerical simulations, demonstrating the improved performance of the proposed scheme compared to an unprotected realization of the 2PQRM. We have considered typical values of experiments involving ${}^{40}{\rm Ca}^{+}$ ions~\cite{Gerritsma10,Gerritsma11}, which have been gathered in Table~\ref{tab:1}, together with the parameters used to model magnetic-field fluctuations (see Appendix~\ref{app:a} for further details).

\begin{table*}[t] 
\begin{center}
    \begin{tabular}{c c c c c c c c c}
    \hline
    $\omega_I$ $\qquad$ & $\nu$ $\qquad$ &  $\Omega_{r,b}$ $\qquad$ & $\eta$ $\qquad$& $\eta_c$ $\qquad$& $\phi_{r,b}$ $\qquad$& $\phi_c$ $\qquad$& $\tau$ $\qquad$& $T_2$ \\
    \hline\hline\\
    $2\pi \times 4\cdot 10^{14}$ Hz $\qquad$ &  $2\pi \times 2$ MHz $\qquad$ & $ 2\pi \times 100$ kHz $\qquad$ & $0.06$ $\qquad$ & $0.01$ $\qquad$ & $\pi$ $\qquad$ & $0$ $\qquad$ & $100\ \mu$s $\qquad$ & $3$ ms\\
    \hline
    \end{tabular}
\end{center}
\caption{Values of the  trapped-ion parameters used in the numerical simulations, together with the coefficients of the magnetic-field fluctuations. The frequency splitting of the trapped-ion is $\omega_I$, $\nu$ defines the trap frequency, while the employed lasers to produce detuned second-order sidebands carry Rabi frequency $\Omega_{r,b}$, Lamb-Dicke parameter $\eta_{r,b}\equiv \eta$ and initial phase $\phi_{r,b}$. The additional carrier interaction is driven with a Rabi frequency $\Omega_c$ that takes a dual role, as discussed in Sec.~\ref{sub:dd} (see third paragraph of Sec.~\ref{sec:NS} for further details), $\eta_c$ is the Lamb-Dicke parameter  and $\phi_c$ the initial laser phase. The magnetic-dephasing noise sets a coherence time $T_2$ and exhibits a relaxation time $\tau$ (see Sec.~\ref{sub:noise}) for further details). Unspecified parameters, such as detunings of the second sidebands, $\delta_{r,b}$, or Rabi frequency $\Omega_c$,  will be different depending on the simulated 2PQRM.}
\label{tab:1}
\end{table*}

We simulate the noisy trapped-ion dynamics described by the Hamiltonian
\begin{align}
\label{eq:TIsim}
&H_{TI,n}^{I}= \frac{\xi(t)}{2}\sigma_z\nonumber\\
&+\sum_{j} \frac{\Omega_j}{2} \left[\sigma^+ e^{i\eta_j(ae^{-i\nu t}+\adag e^{i\nu t} )}e^{i(\omega_I-\omega_j)t-i\phi_j}+{\rm H.c.} \right],
\end{align}
which is accomplished only after the assumption of optical RWA. Note that this approximation is known to hold  in this setup as a consequence of the large frequency $\omega_I$, and $\omega_j\approx \omega_I$. It is worth emphasizing that no further approximations have been taken, hence our simulations take Eq.~(\ref{eq:TIsim}) as the starting point. The previous Hamiltonian is indeed general as it does not specify either the number of employed lasers nor their corresponding parameters. It therefore encompasses the protected and unprotected realization of the 2PQRM, depending on the utilized lasers. Furthermore, as we have seen in Sec.~\ref{sec:TI}, the 2PQRM can be written in a compact way as
\begin{align}
H_{\rm 2PQRM}^{U,P}= \frac{\tilde{\Omega}^{U,P}}{2}\sigma_\parallel^{U,P}+\tilde{\omega}_0^{U,P}\adaga
+\tilde{g}^{U,P}\left[a^2+(\adag)^2 \right]\sigma_\perp^{U,P},
\end{align} 
where the superscript $U$ ($P$) explicitly specifies the unprotected (protected) realization. Although both realizations are completely equivalent, the relation between trapped-ion parameters and the simulated values of the $H_{\rm 2PQRM}$ is not. This can be found in the Table~\ref{tab:2}. In addition, the spin basis of the simulated 2PQRM results in $\sigma_\parallel^{U}=\sigma_z$, $\sigma_\perp^{U}=\sigma_x$ while  $\sigma_\parallel^{P}=\sigma_x$, $\sigma_\perp^{P}=\sigma_z$. 

\begin{table}[t] 
\begin{center}
    \begin{tabular}{c  c  c  c  c}
      \hline\\
     $\tilde{\Omega}^{U}$ $\quad$& $\tilde{g}^{U}$ $\quad$& $\tilde{\omega}_0^{U,P}$ $\quad$& $\tilde{\Omega}^{P}$ $\quad$& $\tilde{g}^{P}$ \\
    \hline\hline\\
    $(\delta_b+\delta_r)/2$ $\quad$&   $\eta^2\Omega/4$ $\quad$&$(\delta_b-\delta_r)/4$ $\quad$& $\Omega_c-\Omega_{DD}$ $\quad$& $\eta^2\Omega/8$\\
    \hline
    \end{tabular}
\end{center}
\caption{Relation between trapped-ion and simulated parameters of the 2PQRM, for both protected and unprotected realization.}
\label{tab:2}
\end{table}

We demonstrate the advantage of the proposed scheme to achieve a 2PQRM, averaging several stochastic trajectories of the simulated noisy trapped-ion Hamiltonian, given in Eq.~(\ref{eq:TIsim}). In particular, we consider three different situations in the USC regime, namely, $\tilde{g}/\tilde{\omega}_0=0.1$, $0.2$ and $0.3$ with $\tilde{\Omega}/\tilde{\omega}_0=3$, $2$ and $1$, respectively. The considered initial states are $\left|\downarrow \right>_\perp\left|0 \right>$ for $\tilde{g}/\tilde{\omega}_0=0.1$, and $\left|\uparrow \right>_\parallel\left|2 \right>$ for the other two. Recall that the spectral collapse takes place at $\tilde{g}/\tilde{\omega}_0=0.5$. The targeted coupling $\tilde{g}/\tilde{\omega}_0=0.1$ leads to $\tilde{\omega}^U_0=2\pi\times 900$ Hz and $\tilde{\omega}^P_0=2\pi\times 450$ Hz. For $\tilde{g}/\tilde{\omega}_0=0.2$ and $0.3$, the corresponding simulated bosonic frequency are even smaller, $\tilde{\omega}^U_0=2\pi\times 450$ Hz and $\tilde{\omega}^U_0=2\pi\times 300$ Hz, respectively. Note that $\tilde{\omega}^P_0=\tilde{\omega}^U_0/2$ as a consequence of the resulting couplings $\tilde{g}^P=\tilde{g}^P/2$. Therefore, the dynamics of the simulated 2PQRM in the USC regime appears in the range of few milliseconds, since $\tilde{g}^U=2\pi\times 90$ Hz. From the previous values, and those in Tables~\ref{tab:1} and~\ref{tab:2}, the rest of the parameters follow, except $\Omega_{DD}$ for the decoupling scheme, which is set to $\Omega_{DD}=2\pi \times 20$ kHz. Hence, we drive the carrier with an intensity $\Omega_c=\Omega_{DD}+\tilde{\Omega}$, which results in $\Omega_c\approx \Omega_{DD}$ as a consequence of the aimed 2PQRM parameters,  $\tilde{\Omega}\leq 3\tilde{\omega}_0^P \leq 2\pi\times 1.35$ kHz.  Note that decoupling is ensured since $\Omega_{DD}> f_{cr}=1/(2\pi\tau)$, while $\Omega_{DD}\ll \nu$ enables a correct driving of detuned second sidebands, $\delta_{r,b}=\Omega_{DD}\mp 2\tilde{\omega}_0^P$, as explained in Sec.~\ref{sub:dd}. It is worth mentioning that the time scale of the protected realization is doubled with respect to the unprotected one, which together with smaller detunings $\delta_{r,b}$, make the unprotected scheme better suited to fulfill all the approximations listed in Sec.~\ref{sec:TI}. Succinctly, if no dephasing noise is included, the trapped-ion simulation of the 2PQRM is better accomplished when no protection scheme is employed, as the latter relies on an additional RWA and required detunings deteriorate the vibrational RWA. In contrast, when realistic dephasing noise is included, dynamical decoupling scheme largely surpasses in performance its unprotected counterpart. This is illustrated in the Appendix~\ref{app:b}, where we show results of noiseless trapped-ion simulation of a 2PQRM with both schemes, i.e. when $\xi(t)\equiv0 \ \forall t$. Accordingly, it can be stated that the reported deterioration of simulated dynamics arises from magnetic-field fluctuations and not from the different RWAs used in the construction of the 2PQRM Hamiltonian.


In Fig.~\ref{fig:1} we show the results, obtained after an ensemble average over $400$ stochastic trajectories of an initial state evolving under $H_{TI,n}^{I}$, Eq.~(\ref{eq:TIsim}). In particular, we show the time evolution of relevant observables, namely bosonic excitations, $\left<\adaga\right>$, and $\left<\sigma_{\parallel} \right>$ or $\left<\sigma_{\perp} \right>$ depending on the considered initial state for each of the three different scenarios, listed above.   For a better comparison, the targeted 2PQRM is displayed as well. It is indeed clear that the protected scheme allows for a faithful realization of the 2PQRM, allowing to explore dynamics otherwise spoiled by magnetic-field fluctuations. Note that $\tilde{\omega}_0^{U}=2\tilde{\omega}_0^{P}$, which leads to longer real-time simulations. In the results shown in Fig.~\ref{fig:1}, the total time of the unprotected realization is $5$ ms, while for the protected scheme reaches $10$ ms.  In order to quantify the deviation of simulated trapped-ion dynamics with respect to the ideal 2PQRM, we make use of the standard fidelity
\begin{align}
F^{U,P}(t)=\left|\left<\psi(t)^{U,P} \right|\left.\psi(t)^{\rm 2PQRM}  \right>\right|
\end{align}
where $\left|\psi(t)^{U,P} \right>$ represents the  evolved state including internal  and motional degrees of freedom, either on the unprotected (U) or protected (P) scheme, while $\left|\psi(t)^{\rm 2PQRM} \right>$ corresponds to the evolved  state under the ideal and targeted 2PQRM Hamiltonian, $H_{\rm 2PQRM}$. An ideal trapped-ion realization would then provide the highest fidelity, $F\equiv 1$.  In Fig.~\ref{fig:2}, and for the sake of a better visualization, we show the time evolution of the infidelity, i.e., $1-F^{U,P}(t)$, for the worst case among the three shown in Fig.~\ref{fig:1}, namely, $\tilde{g}/\tilde{\omega}_0=0.3$, $\tilde{\Omega}/\tilde{\omega}_0=1$ and initial state $\left|\downarrow \right>_\parallel\left|2 \right>$, which correspond to the panels (e) and (f). The enhancement of the protected scheme is noticeable, which in this case leads to $1-F^{P}\approx 10^{-1} (1-F^{U})$.  At the end of the evolution, the fidelities drop to $0.75$ in the unprotected scheme for the three cases, while they remain above $0.97$ for the protected one.

Then, and despite of the prolonged times, trapped-ion simulation of the 2PQRM based on the protected scheme faithfully accomplishes the targeted 2PQRM.
This suggests that an exploration of USC regime dynamics of the 2PQRM is indeed feasible in a trapped-ion setup, even in the presence of magnetic-dephasing noise. Nonetheless, we stress that another sources of noise might distort the reported dynamics, which have not been included  here as they are expected to have an impact in longer time scales, as for example heating noise, whose rate can be estimated as $1$ phonon per $100$ ms. Therefore, the latter may deteriorate the reported results  at final times as the simulated dynamics in the protected case reaches $10$ ms. While heating noise will have an impact in both schemes, the improvement of the protected one  with respect to its unprotected counterpart is noticeable at shorter times (because of the elimination of the dephasing noise) where heating is negligible. In addition, the total evolution time for the protected case can be further reduced by adjusting the considered parameters. Finally, we remark that the proposed scheme can be subject to further optimization depending on specific setup parameters and noise conditions. 

\section{Summary}
\label{sec:conc}

In summary, we have proposed  a continuous dynamical decoupling scheme to attain a two-photon quantum Rabi model in a trapped-ion setup which is shown to be robust against magnetic-field fluctuations. This source of noise is recognized as the main obstacle in the considered trapped-ion experiments for keeping coherent dynamics. Therefore, robust schemes to overcome  noise effects may be of great interest for long-time quantum simulations. In addition, when targeted dynamics occurs in a time scale comparable to coherence time of the trapped-ion setup, robust schemes become  an essential requirement to have access to its experimental exploration. In this regard, the two-photon quantum Rabi model appears as a good example because while it exhibits interesting physics in the ultrastrong coupling regime, its trapped-ion realization gets necessarily slow and a proper exploration of this regime demands a scheme intrinsically robust against noise.

With help of a continuous dynamical decoupling method, we propose a scheme that can be implemented with current trapped-ion technologies where magnetic-dephasing noise is averaged out. In particular, 2PQRM is accomplished by three lasers, driving carrier, and  second red- and blue-sidebands, respectively, where the former plays a crucial (double) role since permits decoupling from magnetic-dephasing noise and defines the simulated free-energy term of the two-level system. We demonstrate the advantage of such a scheme with respect to an unprotected realization by means of detailed numerical simulations with typical trapped-ion parameters.

Although the applicability of the proposed scheme has been asserted in a trapped-ion setup, it might be employed as well in distinct experimental platforms as superconducting circuits.

\begin{acknowledgments}
This work was supported by the ERC Synergy grant BioQ, the EU STREP project EQUAM and the CRC TRR21. This work was performed on the computational resource bwUniCluster funded by the Ministry of Science, Research and the Arts Baden-W\"urttemberg and the Universities of the State of Baden-W\"urttemberg, Germany, within the framework program bwHPC. J.~C. acknowledges Universit\"at Ulm for a Forschungsbonus.
\end{acknowledgments}

\appendix

\section{Orstein-Uhlenbeck noise to model magnetic-field fluctuations}
\label{app:a}
\begin{figure}
\includegraphics[width=1.2\linewidth,angle=-90]{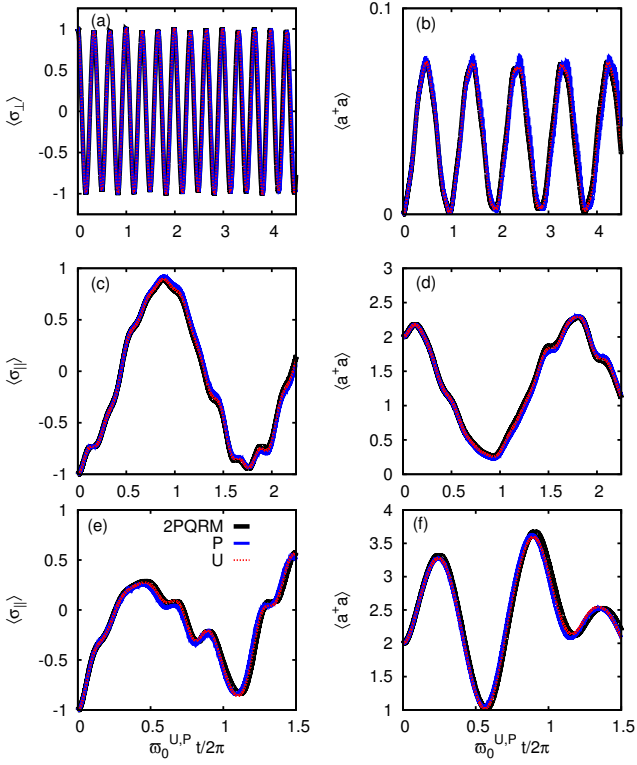}
\caption{(Color online) Dynamics of the noiseless trapped-ion simulation of the 2PQRM, with protection (P) (blue solid line) and without (U) (red dotted line), together with the ideal  2PQRM (black solid line). In the left column qubit observables, and $\left<\adaga\right>$ in the right. The trapped-ion results were obtained from a single evolution under the Hamiltonian Eq.~(\ref{eq:TIsim}) since $\xi(t)\equiv 0$. Three different scenarios are examined. Upper panels, (a) and (b), correspond to $\tilde{g}/\tilde{\omega}_0=0.1$, $\tilde{\Omega}/\tilde{\omega}_0=3$ and initial state $\left|\downarrow \right>_\perp\left|0 \right>$. Middle row, (c) and (d), $\tilde{g}/\tilde{\omega}_0=0.2$, $\tilde{\Omega}/\tilde{\omega}_0=2$ and initial state $\left|\uparrow \right>_\parallel\left|2 \right>$, and for the bottom panels, (e) and (f), $\tilde{g}/\tilde{\omega}_0=0.3$, $\tilde{\Omega}/\tilde{\omega}_0=1$ and initial state $\left|\downarrow \right>_\parallel\left|2 \right>$. In this noiseless scenario the three curves overlap, in contrast to the results shown in Fig.~\ref{fig:1}. The total time for the unprotected is  $5$ ms, while it is doubled for the protected scheme, $10$ ms. See Tables~\ref{tab:1} and~\ref{tab:2}, as well as the main text, for further details.}
\label{fig:3}
\end{figure}
In this Appendix we provide additional material regarding how to model magnetic-field fluctuations by an Orstein-Uhlenbeck (OU) noise~\cite{Uhlenbeck30,Wang45,Gillespie96,Gillespie96bis}. A Markov process $\xi$ evolving in time according to Eq.~(\ref{eq:OU}) is known as OU noise.  As discussed in the main text, this Gaussian noise depends on two variables, namely correlation or relaxation time $\tau$ and diffusion constant $c$. In this context, an important property is the so-called spectral density $S(f)\equiv \lim_{T\rightarrow \infty} 2|\hat{\xi}(f)|^2/T$ which quantifies the portion of noise intensity at a particular frequency, with $\xi(t)=\int df \hat{\xi}(f)e^{-2\pi i f t}$ the Fourier transform. It is then defined in terms of the auto-covariance $C(t')$ of a stationary $\xi$, which for an OU process can be calculated~\cite{Gillespie96,Gillespie96bis}
\begin{align}
C(t')\equiv\overline{\xi(t)\xi(t+t')}=\frac{c\tau}{2}e^{-t'/\tau} \qquad t'\geq 0.
\end{align}
Then, it can be shown that $S(f)$ can be written as
\begin{align}
S(f)&=4\int_0^\infty dt'\  C(t')\cos(2\pi f t') \nonumber\\&=\frac{2c\tau^2}{1+4\pi^2\tau f^2} \qquad f\geq 0.
\end{align}
Additionally, the total intensity of the noise is given by $\overline{\xi^2(t)}=C(0)=c\tau/2$ or equivalently obtained from $\int_0^\infty df S(f)$. Note that it is useful to define a characteristic frequency, $f_{cr}=1/(2\pi\tau)$, which settles the transition from white noise, $S(f<f_{cr})\propto f^0$ to Brownian noise, $S(f>f_{cr})\propto f^{-2}$. Having in mind the relation between $\tau$ and $c$ and the noise properties, we consider now how they relate to the loss of coherence and $T_2$ to model magnetic-field fluctuations.

Let consider a simple case where an initial state $\upx$, $\sigma_x\upx=+\upx$, evolves under a noisy Hamiltonian, $H=\xi(t) \sigma_z/2$ that effectively models magnetic-field fluctuations. Therefore, the expectation value of $\sigma_x$ after a single run and time $t$ is given by
\begin{align}
\left<\sigma_x(t) \right>=\cos\left(\Xi(t)\right)
\end{align}
with $\Xi(t)=\int_0^tdt'\ \xi(t)$ the integral of the stochastic noise. After stochastic average, the result becomes
\begin{align}
\overline{\left<\sigma_x(t) \right>}=\overline{\cos\left(\Xi(t)\right)}=e^{-\frac{1}{2}\overline{\Xi^2(t)}},
\end{align}
where we have assumed that the stochastic noise $\xi$ is Gaussian, that is, the moments fulfill $\overline{\xi^{2n}}=\overline{\xi^{2}}^n (2n)!/(2^n n!)$, as is the case for an OU noise. Finally, $\overline{\Xi^2(t)}$ can be obtained analytically as a function of $\tau$ and $c$~\cite{Gillespie96}; thus, we obtain a relation between $c$, $\tau$ and $T_2$ since $\overline{\left<\sigma_x(T_2) \right>}=e^{-1}$, which leads to
\begin{align}
c=\frac{2}{\tau^2\left(T_2-\tau\left(\frac{3}{2}-2e^{-T_2/\tau}+\frac{1}{2}e^{-2T_2/\tau} \right) \right)}.
\end{align}
Therefore, knowing $T_2$ and a $\tau$ the corresponding diffusion constant is determined. In the context of trapped ions, the noise is short correlated, i.e., $\tau\ll T_2$, which simplifies the previous equation to $c\approx 2/(T_2 \tau^2)$. This correctly predicts the exponential decay of $\overline{\left<\sigma_x(t) \right>}$ observed experimentally~\cite{Wineland98}.

\section{Noiseless trapped-ion Hamiltonian}
\label{app:b}
Here we show that, based on the developments discussed in the main text, the performance of the unprotected trapped-ion scheme in fact surpasses its protected counterpart  if magnetic-field fluctuations are not included in the model. As commented in the main text, in this noiseless scenario, the proposed dynamical decoupling scheme is expected to provide worse results than its unprotected counterpart as the former relies on an additional rotating-wave approximation (see Sec.~\ref{sub:dd}) and the required detunings for the second red- and blue-sidebands are higher, thus, they may deteriorate the vibrational RWA.

We simulate the general trapped-ion Hamiltonian, Eq.~(\ref{eq:TIsim}), assuming only the optical RWA (as in Sec.~\ref{sec:NS}) and neglecting any stochastic fluctuation, i.e., $\xi(t)\equiv 0\ \forall t$. Therefore, no stochastic average is needed. The simulated 2PQRM parameters, as well as the initial states, are the same as the ones presented in Sec.~\ref{sec:NS} (Fig.~\ref{fig:1}), which facilitate a direct comparison. In the Fig.~\ref{fig:3}, we show the results of these noiseless simulations. Note that the simulated results lie almost perfectly on top of the ideal 2PQRM. However, a closer inspection of the fidelities reveals the better performance of the unprotected scheme in the noiseless scenario: while the results of the protected scheme do not differ significantly from those obtained when the noise is included, $F^{P}\gtrsim 0.98$, the unprotected scheme reaches higher fidelities, $F^{U}\gtrsim 0.99$. We recall that $F^{U}$ drops to $0.75$ when noise is incorporated (see Sec.~\ref{sec:NS}). Hence, the deterioration  of the  simulated 2PQRM without protection in the noisy scenario is an exclusive consequence of the magnetic-field fluctuations (see Fig.~\ref{fig:1}), which can be suppressed using the protected scheme. This supports the suitability of the discussed continuous dynamical decoupling scheme to enhance the realization of the 2PQRM under realistic magnetic-field fluctuations, as discussed in the main text.

\end{document}